\documentclass[%
reprint,
superscriptaddress,
amsmath,amssymb,
aps,
prl,
longbibliography
]{revtex4-2}

\usepackage{graphicx}
\graphicspath{ {Figures/} }
\usepackage{comment}

\usepackage{dcolumn}
\usepackage{bm}
\usepackage[export]{adjustbox}
\usepackage[normalem]{ulem}
\usepackage{mathtools} 

\DeclarePairedDelimiter\ket{\lvert}{\rangle}
\DeclarePairedDelimiterX\braket[2]{\langle}{\rangle}{#1 \delimsize\vert #2}

\usepackage[utf8]{inputenc}
\usepackage[T1]{fontenc}

\usepackage{mathrsfs}
\usepackage{dsfont}

\usepackage{amsmath} 
\usepackage{amsthm}  
\usepackage{amsfonts}
\usepackage{amsthm}

\usepackage{physics}

\usepackage{hyperref}
\hypersetup{pdfpagemode=UseNone}

\usepackage{orcidlink}

\newtheorem{theorem}{Theorem}
\newtheorem{definition}[theorem]{Definition}

\newcounter{resul}
\newtheorem{result}[resul]{Result}

\newcounter{rem}
\newcommand{\mc}[1]{\mathcal{#1}}

\newcommand{\beq}{\begin{equation}}
\newcommand{\eeq}{\end{equation}}

\def\>{\rangle}
\def\<{\langle}
\newcommand{\proj}[1]{| #1 \rangle\! \langle #1 |}

\def\tr{{\rm Tr}}

\def\textbf#1{{\bf #1}}
\newcommand{\Cx}{\mathbb{C}}

\newcommand{\Rl}{\mathds{R}}
\gdef\lvert{\delimiter"426A30C }

\def\<{\langle}

\newcommand{\deff}{d_{\mathrm{eff}}}

\usepackage{xcolor}

\begin{document}

\title{When Isolated Quantum Systems Appear Classical
}
\author{Thiago R. de Oliveira}
\affiliation{Instituto de Física, Universidade Federal Fluminense, Av. Litorânea s/n, Gragoatá 24210-346, Niterói, RJ, Brazil}
\author{Pedro S. Correia}
\affiliation{Departamento de Ciências Exatas, Universidade Estadual de Santa Cruz, Ilhéus, Bahia 45662-900, Brazil}
\author{Tiago Debarba}
\affiliation{Departamento Acad{\^ e}mico de Ci{\^ e}ncias da Natureza - Universidade Tecnol{\'o}gica Federal do Paran{\'a}, Campus Corn{\'e}lio Proc{\'o}pio - Paran{\'a} -  86300-000 - Brazil.}
\author{Gabriel Dias Carvalho}
\affiliation{Física de Materiais, Universidade de Pernambuco, 50720-001, Recife, PE, Brazil}
\author{Ra\'ul O. Vallejos}
\affiliation{Centro Brasileiro de Pesquisas F\'{\i}sicas, Rua Dr. Xavier Sigaud, 150, Rio de Janeiro, RJ, Brazil}
\author{Fernando de Melo}
\email{fmelo@cbpf.br}
\affiliation{Centro Brasileiro de Pesquisas F\'{\i}sicas, Rua Dr. Xavier Sigaud, 150, Rio de Janeiro, RJ, Brazil}%
\date{\today}

\begin{abstract}
The emergence of classical behavior and the origin of thermal equilibrium are two central problems in the foundations of physics. In the standard accounts, both phenomena are routinely explained through interactions with an external environment: decoherence suppresses quantum interference, while coupling to an external environment drives relaxation toward equilibrium. Over the last decades, however, it has become clear that equilibration and thermalization can arise even in fully isolated quantum systems, in the operational sense that the expectation values of relevant observables remain close to equilibrium values for most of the time. Here, we ask whether the same intrinsic equilibration mechanism can also account for the emergence of classical behavior.  Extending bounds from the theory of equilibration in closed systems, we derive sufficient conditions under which a time-evolved pure state becomes, for most times, operationally indistinguishable from a classical mixture associated with a chosen physical property. We identify two complementary routes to such operational classicality: either the chosen property almost commutes with the system Hamiltonian, or the observables used to probe the system lose access to the remaining coherence after equilibration. Our results show that classical behavior need not be confined to the energy basis and may emerge even when substantial coherence remains present in the equilibrium state. This establishes a direct connection between two foundational questions: the origin of equilibration in isolated quantum systems and the quantum-to-classical transition.
\end{abstract}
\maketitle

\noindent \textit{Introduction:} Understanding how macroscopic behavior emerges from microscopic laws is a central goal of modern physics. Two prominent examples are the emergence of thermodynamic behavior from reversible microscopic dynamics, and the emergence of classical behavior from quantum mechanics. Although both problems concern the relationship between microscopic and macroscopic descriptions, they are traditionally addressed within different frameworks.

Historically, the orthodox interpretation addressed the quantum-to-classical transition by postulating a distinction between quantum and classical domains \cite{bohr1996discussion}. While successful as a practical framework, this viewpoint leaves open the question of how classical behavior emerges from an underlying quantum description and, therefore, appears difficult to reconcile with a fully reductionist view of nature. The tension is particularly evident in thought experiments such as Schrödinger's cat \cite{schrodinger1935gegenwartige}, where quantum mechanics allows superpositions of macroscopically distinct states that are never observed in practice. Explaining the apparent absence of such macroscopic quantum behavior remains one of the central challenges in the foundations of physics.

The standard account of the quantum-to-classical transition is provided by decoherence theory \cite{decoherence_RMP,zurekRMP}. In this framework, interactions with an external environment suppress quantum interference in a preferred basis, rendering the system effectively classical with respect to appropriate observables. Similarly, traditional approaches to thermalization often invoke a thermal bath whose interaction with the system drives relaxation toward equilibrium. In both cases, emergent macroscopic behavior is explained through coupling to external degrees of freedom.

Over the last decades, however, it has become clear that equilibration and thermalization can arise even in completely isolated quantum systems. Although the global state continues to evolve unitarily and remains pure, the expectation values of physically relevant observables spend most of their time close to stationary values, with only small fluctuations. In this operational sense, the evolving state becomes effectively indistinguishable from its equilibrium description, despite the absence of any fundamental loss of information. This operational notion of equilibration is now supported by rigorous theoretical results~\cite{gogolin2016,Mori2018} and a growing body of experimental evidence~\cite{Langen2015,Ueda2020}.

These developments naturally raise a broader question. If equilibration can account for the emergence of thermodynamic behavior without invoking an external environment, can the same mechanism also explain the emergence of classical behavior? Equilibration already implies a form of operational decoherence in the energy basis, since the time-evolved state becomes indistinguishable from its energy-dephased equilibrium state for a restricted class of observables \cite{Reimann08, Linden08}. The central question addressed in this work is whether this mechanism can induce operational classicality for other physical properties.


We extend the main bound from the theory of equilibration of closed systems \cite{Reimann08, Linden08}, to obtain sufficient conditions for a quantum property to be operationally classical.
Our perspective is explicitly operational: rather than requiring the suppression of all quantum coherence, we ask whether it remains detectable through experimentally accessible observables. Indeed, in realistic macroscopic systems one typically has access only to a restricted set of coarse-grained observables measured with finite precision. Operational indistinguishability with respect to this restricted set, therefore, provides a physically meaningful notion of classicality.

We identify two complementary mechanisms through which operational classicality can emerge: either we restrict to properties which almost commute with the system's Hamiltonian, or the observables used to probe the system lose access to the remaining coherence after equilibration. These results establish a direct connection between equilibration and the emergence of operational classicality, and suggest a common dynamical origin for aspects of both thermal and classical macroscopic behavior.
\newline

\noindent\textit{Decoherence:} Let us briefly review the basic idea of decoherence. Consider a system prepared in a pure state that is a coherent superposition in some basis, $|\psi\> = \sum_n c_n^b \,|b_n\>,$
where $\{|b_n\>\}$ are eigenvectors of an observable 
\beq
\hat{B}=\sum_n b_n \,|b_n\>\< b_n| = \sum_n b_n \,\hat{\Pi}_n^B,
\eeq
with $b_n\in\Rl$ and $c_n^b \in \Cx$, such that $\sum_n |c_n^b|^2=1$.
In general, the state $|\psi\>$ is not classical with respect to $B$, since it does not assign a definite value to $\hat{B}$. For instance, in the spin state $|\!\uparrow\>+|\!\downarrow\>$, the observable $\hat{S}^z$ is not well defined prior to measurement. The corresponding ``classical'' (fully decohered) state in the $\hat{B}$-basis is obtained by removing the off-diagonal coherence,
\beq
\rho_B \,=\, \sum_n |c_n^b|^2 \,|b_n\>\< b_n|
\,=\, \sum_n \hat{\Pi}_n^B \,\proj{\psi}\, \hat{\Pi}_n^B
\,\equiv\, \mathcal{C}_B(\proj{\psi}),
\eeq
with $\mc{C}_B$ the dephasing channel in the eigenbasis of $\hat{B}$.

The state $\rho_B$ contains no coherence in the eigenbasis of $\hat{B}$ and may, therefore, be interpreted as describing a classical statistical uncertainty about the value of the physical property $\hat{B}$: the property $\hat{B}$ is well-defined in $\rho_B$, albeit unknown.

A central goal of decoherence theory is to explain how a state can become operationally indistinguishable from $\rho_B$ without invoking a fundamental collapse postulate. In the standard scenario, this behavior arises through interactions with an external environment. The joint evolution correlates system states with distinct environment states,
\beq
|\psi\rangle\otimes|e_0\rangle
\rightarrow
\sum_n c_n |b_n\rangle\otimes|e_n\rangle.
\eeq
If the states of the environment become approximately orthogonal, tracing out the environment yields a reduced state close to $\rho_B$.

Since a quantum state is coherent in many different bases, it is essential to identify the basis in which classical behavior emerges. In decoherence theory, the system–environment interaction selects a preferred (pointer) basis \cite{zurek2006}. In simple models \footnote{this is the simple case where $H_{int}$ couples only a single system observable to the environment. In the case of multiple observables coupled to the environment, Zurek proposed a measure, the predictability sieve, to identify the pointer states.}, this corresponds to observables whose projectors approximately commute with the interaction Hamiltonian,
 $  [\hat{\Pi}_n^B,\hat{H}_{\rm int}] \approx0.$

Importantly, the coherence is not destroyed globally: the combined system–environment state remains pure and continues to evolve unitarily. Classicality emerges because the coherence becomes inaccessible when observations are restricted to the system alone.

From an operational perspective, decoherence can be understood in terms of distinguishability. Classical behavior emerges whenever the observables available to an observer are unable to distinguish the coherent state $|\psi\rangle\langle\psi|$ from its dephased counterpart $\rho_B$. This viewpoint shifts the focus from the state alone to the interplay between coherence and the observables used to probe it, a perspective that will be central to our discussion of equilibration-induced classicality.
\newline

\noindent\textit{Equilibration:} As in quantum mechanics, statistical mechanics still faces foundational questions that remain actively debated, such as the origin of the second law of thermodynamics. A closely related question is why a macroscopic system, if left unperturbed, should equilibrate and thermalize. A traditional approach is to couple the system to an environment (a thermal bath) and show that, in suitable regimes, this interaction drives the system toward an equilibrium thermal state.

In both the decoherence program and the open-system approaches to thermalization, one explains emergent macroscopic behavior by appealing to an external environment and to the limit of long times and large baths. At the same time, one should keep in mind that the system itself is macroscopic, and one does not have full access to its density matrix but only to a small set of coarse-grained observables, measured with finite precision. We, therefore, adopt an operational viewpoint: it is sufficient to show that such macroscopic observables exhibit equilibrium (thermal) behavior \footnote{This perspective already appears, for instance, in Kinchin's approach to the foundations of statistical mechanics~\cite{Khinchin1949}}.

Over the last decades, a complementary line of work has shown that equilibration and thermalization can occur statistically even when the system is completely isolated, provided that a limited set of observables is available. Let us briefly review the equilibration framework for closed quantum systems.

Consider an isolated system prepared in an initial pure state $|\psi\>=\sum_n c_n^H |E_n\>$, with Hamiltonian $\hat{H}=\sum_n E_n |E_n\>\<E_n| = \sum_n E_n \hat{\Pi}_n^H$. Under unitary evolution, the state $|\psi(t)\>$ remains pure and never converges to a stationary state. Nevertheless, if we only have access to an observable $\hat{O}$, its expectation value $O(t)=\<\psi(t)|\hat{O}|\psi(t)\> = \tr[\hat{O}\,\rho_t]
$ may, after a transient, spend most of the time very close to an equilibrium value $\overline{O}$ with small fluctuations. In this situation, we say that the system equilibrates with respect to $\hat{O}$. 
The quantity $\overline{O}$ plays the role of an equilibrium value, and one can associate an equilibrium state $\rho_H$ such that $\overline{O}=\tr(\hat{O}\,\rho_H)$. 
To formalize this notion, we define the infinite-time average of a function $f$ by
\beq
\overline{f}\,\equiv\,\lim_{T\to\infty}\frac{1}{T}\int_0^T dt\, f(t).
\eeq
Under very general conditions~\footnote{The modern results were obtained in \cite{Reimann08, Linden08}. Later, it was noticed that von Neumann had already derived closely related statements. An important earlier reference is \cite{Tasaki98}.} one can bound the time-averaged fluctuations as
\beq
\overline{(\Delta O)^2} \equiv \overline{\big( O(t)-\overline{O}\big)^2}\;\leq\; \frac{\|\hat{O}\|_\infty^2}{d_{\rm eff}},
\eeq
where $d_{\rm eff}=1/\tr[\rho_H^2]$ is the effective dimension. Thus, if $\|\hat{O}\|_\infty\ll d_{\rm eff}$ the observable $\hat{O}$ equilibrates. Hereafter, the symbol $\|\cdot \|_\infty$ will denote the operator norm.

For Hamiltonians with nondegenerate energy gaps, the infinite-time averaged state is simply the dephased initial state in the energy eigenbasis,
\beq
\rho_H = \sum_n |c_n|^2 |E_n\>\<E_n| = \sum_n \hat{\Pi}_n^H\,\rho_0\,\hat{\Pi}_n^H \equiv \mathcal{C}_H(\rho_0),
\label{Eq. rho_average}
\eeq
where $\mathcal{C}_H$ denotes the dephasing channel in the $H$ basis.

For local many-body Hamiltonians, the many-body level spacing typically decreases exponentially with system size. Consequently, physically realistic initial states are often expected to populate exponentially many energy eigenstates, leading to an effective dimension that grows rapidly with the number $N$ of subsystems,  and is frequently exponential in system size \footnote{Note that even if we restrict the initial state to an energy shell of macroscopic size, we still would get an $\deff \sim \exp(N)$. See Reimann's argument.}. Correspondingly, one expects $\deff\sim \exp(N)$. In this regime, equilibration holds for a broad class of observables whose norms grow at most polynomially with $N$ (for instance, local observables).

Although Eq.~\eqref{Eq. rho_average} is stated for Hamiltonians with nondegenerate energy gaps, analogous equilibration results hold under considerably more general conditions, including finite-time windows, infinite-dimensional systems, and broad classes of time-dependent Hamiltonians \cite{Short2012,Reimann2012,Passos25}.

Operationally, for a macroscopic system probed only through $O$, it becomes impossible to distinguish the evolving pure state from the highly mixed state $\rho_H$ for most times; in this sense, the system equilibrates for all practical purposes. Importantly, the global state never converges to $\rho_H$: equilibration is observable-dependent and probabilistic, in the sense that $O(t)$ continues to evolve but exhibits only small fluctuations for typical times.

This perspective suggests an interpretation of equilibration as an effective dephasing mechanism acting on observables rather than states. The equilibrium expectation value depends only on the component of $\hat{O}$ that is diagonal in the energy basis, while the off-diagonal components become operationally inaccessible through temporal averaging. 
In this sense, equilibration implements the analog of a dephasing channel in the Hamiltonian basis at the level of observable predictions.
\newline

\noindent\textit{Decoherence and Equilibration:} As discussed above, equilibration in isolated quantum systems implies that, for a restricted set of observables, the time-evolved state $\rho_t$ becomes, on average, operationally indistinguishable from the equilibrium state $\rho_H$. Since $\rho_H$ is diagonal in the energy eigenbasis, this can be interpreted as a form of decoherence in the Hamiltonian basis, as witnessed by the accessible observables. This observation naturally raises the question of whether equilibration can induce an analogous suppression of observable coherence for physical properties other than energy.

From an operational perspective, the relevant question is whether the evolving state $\rho(t)$ can be distinguished from the classical state $\rho_B$ associated with a physical property $\hat{B}$ using a restricted family of observables. In other words, we ask whether there exists a state diagonal in the $\hat{B}$ basis such that, for most times,
\begin{equation}
\mathrm{Tr}[\hat{O} \rho_t] \approx \mathrm{Tr}[\hat{O} \rho_B]
\end{equation}
for all observables $\hat{O}$ in the set of accessible physical properties. This provides a natural notion of classicality that does not rely on coupling to an external environment.

Notice that we do not expect operational classicality to emerge for arbitrary physical properties. Rather, it should depend on the interplay among the property under consideration, the system dynamics, and the set of available observables for probing the system.
The results below identify sufficient conditions under which equilibration induces operational classicality for physical properties other than energy. In particular, classical behavior need not be tied exclusively to the Hamiltonian basis, but may emerge whenever the coherence associated with a property becomes inaccessible to the observables relevant for probing the system after equilibration. 
\newline

\noindent\textit{Results:} 
To formalize the ideas above, we introduce the following definitions:
\begin{definition}[\textbf{Operational indistinguishability}]
Given a set $\mathcal{M}$ of experimentally accessible observables, suppose that the expectation value of each $O\in\mathcal M$ can only be determined up to an experimental imprecision $\epsilon$, so that a measurement on a state $\rho$ yields $\tr(O\rho)\pm\epsilon$. We say that two states $\rho$ and $\sigma$ are operationally indistinguishable with respect to $\mathcal  M$ and precision $\epsilon$ if, for every observable $O\in\mathcal M$,
\[
\left|
\tr\left[O(\rho-\sigma)\right]
\right|
\leq 2\epsilon.
\]
\end{definition}

Indeed, under the above condition, the experimental intervals $\tr(O\rho)\pm\epsilon$ and $\tr(O\sigma)\pm\epsilon$ overlap for every accessible observable. Hence, no measurement available within $\mathcal M$ can resolve the difference between $\rho$ and $\sigma$ at the assumed experimental precision.

\begin{definition}[\textbf{Operational classicality of a physical property}] Given an imprecision $\epsilon$, a property  $\hat{B}$ is deemed operationally classical for a given system if its state is operationally indistinguishable from some state diagonal in the $\hat{B}$ basis.
\end{definition}

Finally, notation-wise, for any operator $A$, we denote by
\beq\label{eq:coherence_A}
\hat{A}_B^{\mathrm{coh}} \equiv \hat{A}- \mathcal{C}_B(\hat{A}),
\eeq
the coherence component of $\hat{A}$ in the $\hat{B}$ basis.
The quantities $\|\hat{A}_B^{\mathrm{coh}}\|_1$ and $\|\hat{A}_B^{\mathrm{coh}}\|_\infty$ provide  natural measures of the coherence of $\hat{A}$ with respect to the physical property $\hat{B}$~\cite{baumgratz2014quantifying,streltsov2017colloquium} -- hereafter $\|\cdot\|_1$ denotes the trace norm.
\\
\\

Equilibration can fail to produce operational classicality for two distinct reasons. First, the equilibrium state itself may retain coherence in the $\hat{B}$ basis. Second, even if such coherence is present, the observables used to probe the system may remain sensitive to it. The following two results quantify these two mechanisms separately.

\begin{result}[\textbf{State-based operational classicality}]
\label{result:1} Consider an isolated quantum system with Hamiltonian $\hat{H}$ satisfying the nondegenerate energy-gap condition, and let the initial state have effective dimension $\deff$. Denote by $\rho_H$ the equilibrium state and define $\rho_B := C_B(\rho_H)$, where $C_B$ is the dephasing map in the eigenbasis of the observable $\hat{B}$. Then, for any observable $\hat{O}$ in a given measurement set $\mathcal{M}$,
\begin{equation}
\label{eq:bound1}
\overline{
\left|
\Tr\!\left[\hat{O}(\rho_t-\rho_B)\right]
\right|^2}
\le
\frac{\|\hat{O}\|_\infty^2}{\deff}
+
\|\hat{O}\|_\infty^2
\left\|\rho_{H,B}^{\mathrm{coh}}\right\|_1^2.
\end{equation}
\end{result}

Consequently, if, for every $\hat{O}\in\mathcal{M}$, the above upper bound is smaller than $2\epsilon$, with $\epsilon$ the experimental imprecision, then, for most times, the evolving state $\rho_t$ is operationally indistinguishable from the state $\rho_B$ in which the property $B$ is well defined. Property $B$ is thus operationally classical. Note that we use the squared deviation here, but not in Definition 1. They can be connected by the inequality $\overline{ x }^2\le \overline{ x^2}$.

Result~\ref{result:1} provides a state-based route to operational classicality. The first contribution on the right-hand side of~\eqref{eq:bound1} is the usual equilibration term: it becomes small whenever the effective dimension is large compared with the relevant observational scale, namely when
\(
\|\hat O\|_\infty^2 \ll d_{\mathrm{eff}} .
\)
However, equilibration alone is not sufficient to guarantee the classicality of the property $B$. One must also require the equilibrium state to have little coherence in the eigenbasis of $\hat B$. This is quantified by
\(
\rho^{\mathrm{coh}}_{H,B}:=\rho_H-C_B(\rho_H),
\)
whose trace norm measures the residual coherence of $\rho_H$ in the $\hat B$ basis. Thus, for an observable $\hat O$, the relevant coherence contribution is controlled by
\(
\|\hat O\|_\infty
\left\|\rho^{\mathrm{coh}}_{H,B}\right\|_1 .
\)
Whenever both the equilibration contribution and this coherence contribution are below the experimental resolution, the evolving state $\rho_t$ is, for most times, operationally indistinguishable from the dephased equilibrium state $\rho_B=C_B(\rho_H)$. In this sense,  property $B$ behaves as a classical property relative to the set of accessible observables.


From the above result, one can infer that operational classicality favors properties whose eigenbases are approximately aligned with the eigenbasis of the equilibrium state, so that the residual coherence of $\rho_H$ in that basis is small. This occurs, in particular, when the eigenvectors of $\hat B$ are approximately aligned with those of $\hat H$, implying $\|[\hat B,\hat H]\|_\infty\approx 0$. Such properties play a role analogous to that of the pointer basis in decoherence theory: coherences in this basis become irrelevant to the accessible observables. The analogy is nevertheless not exact. In standard decoherence, the preferred property is selected by the interaction between a system and its environment. Here, by contrast, the mechanism is intrinsic to the isolated system and is determined by the relation between the physical property $\hat B$ and the Hamiltonian $\hat H$. Equilibration, therefore, reproduces some of the key operational features of decoherence without invoking an external bath.

Result~\ref{result:1} attributes classicality to the suppression of coherence in the equilibrium state itself. This requirement, however, may be unnecessarily restrictive. In many situations, substantial coherence can persist in the equilibrium state without observable consequences because the experimentally accessible observables are insensitive to it. This motivates an observable-based notion of classicality.

\begin{result}[\textbf{Observable-based operational classicality}]\label{result:2}
Consider the same assumptions and definitions as for Result~\ref{result:1}.  Then, for any observable $\hat O$ in a given measurement set $\mathcal M$,
\begin{equation}
\overline{
\left|
\Tr\!\left[\hat O(\rho_t-\rho_B)\right]
\right|^2} \le
\frac{\|\hat O\|_\infty^2}{d_{\mathrm{eff}}}
+
\left\|
C_H\!\left(\hat O^{\mathrm{coh}}_B\right)
\right\|_\infty^2
,
\end{equation}
where $C_H$ is the dephasing map in the energy eigenbasis.
\end{result}

As before, suppose that expectation values can only be resolved experimentally up to an imprecision $\epsilon$. If, the above upper bound is smaller than $2\epsilon$ for every $\hat O\in\mathcal M$, then the evolving state $\rho_t$ is, for most times, operationally indistinguishable from the state $\rho_B$ in which the property $B$ is well defined. In this sense,  property $B$ is operationally classical.

Result~\ref{result:2} characterizes classicality at the level of observables rather than states. The quantity $\|\mathcal{C}_H(\hat{O}_B^{\rm coh})\|_\infty^2$ measures the component of the observable that remains sensitive to coherence in the $\hat{B}$ basis after equilibration. Consequently, the equilibrium state may retain substantial coherence while all physically relevant observables become effectively blind to it. Classical behavior, therefore, emerges not necessarily because coherence disappears from the state, but because the remaining coherence becomes operationally inaccessible. This is analogous to equilibration itself: the global state remains pure and continues to evolve unitarily, yet the  accessible observables  become insensitive to this underlying coherence.

\begin{figure}
    \centering
    \includegraphics[width=1\linewidth]{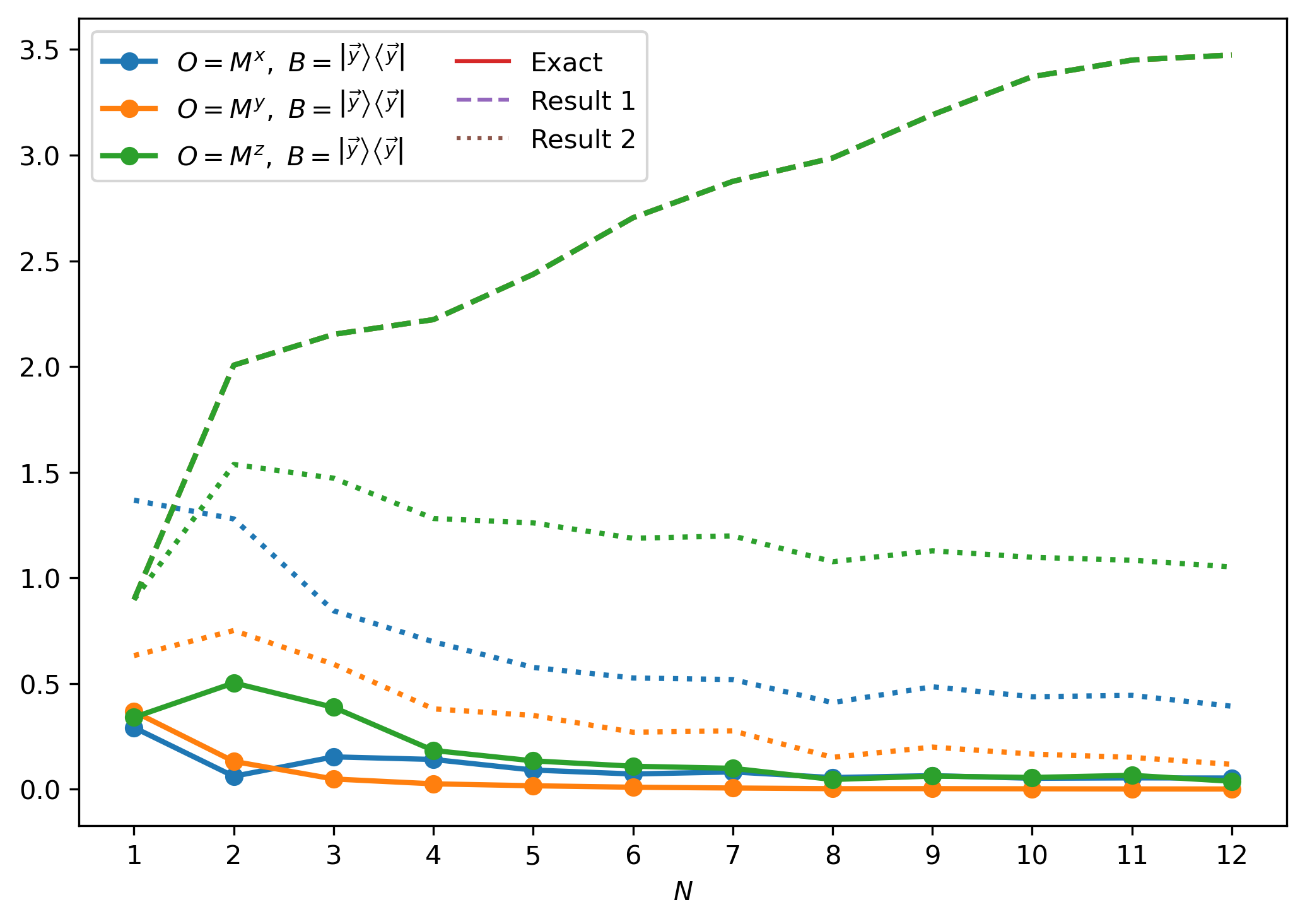}
    \caption{We consider an Ising chain of $N$ spins with both transverse and longitudinal fields and by the following Hamiltonian $ H= \sum_{i=1}^{N-1} J \sigma_i^z \sigma_{i+1}^z + \sum_{i=1}^N  h \; \sigma_i^x + g \; \sigma_i^z$, where $\sigma_i^\alpha$ is the Pauli operator in the direction $\alpha$ acting on the spin on site $i$. The constants have been chosen as $h = 0.5$, $ g = 0.3$,  and $J = 1$ and $M^\alpha = \sum_{i=1}^N \sigma_i^\alpha/N$. We thus consider as a set of accessible observables the magnetization in $x$, $y$, and $z$ directions, and employ them to asses the classicality of physical properties $B$ that are diagonal in the computational basis $\{\ket{\vec{y}}=\ket{y_1, y_2,\ldots,y_N}\}$ in $y$ direction.  The solid curve shows the exact time-averaged fluctuations $\overline{\Delta O^2}$, while the dashed and dotted curves show the bounds from Results 1 and 2, respectively. The three dashed curves corresponding to different observables coincide because the bounds depend only on $||O||_\infty$, which is the same for these observables \label{fig:bounds}. Similar results are also obtained for properties $B$ in the $x$ and $z$ directions.}
\end{figure}

The two results probe complementary notions of coherence. Result~\ref{result:1} concerns the residual coherence of the equilibrium state in the eigenbasis of $\hat B$, whereas Result~\ref{result:2} concerns the sensitivity of the accessible observables to coherences in that basis. More precisely, the second result depends on the component of $\hat O$ that is coherent in the $\hat B$ basis and survives equilibration. This is generally a weaker, measurement-dependent requirement, and therefore provides a broader route to operational classicality. In this way, equilibration can make a property $B$ operationally classical not only when its eigenbasis is close to the energy basis, but also when the coherences associated with that basis are effectively invisible to the accessible observables.  See Fig.~\ref{fig:bounds} for a concrete example.

We may interpret equilibration as a coherence filter: equilibration suppresses the contribution of coherence components that are not protected by conservation laws or dynamical constraints, while protected components may remain observable. Although conservation laws may constrain equilibration itself, for generic many-body systems with relatively few conservation laws, one may expect the coherence that survives equilibration to be strongly constrained for broad classes of observables. Moreover, equilibration is known to occur under very broad and experimentally realistic conditions \cite{Reimann08}, suggesting that operational classicality may likewise emerge quite generally. However, equilibration alone does not guarantee operational classicality: the coherence associated with the physical property under consideration must either be suppressed in the equilibrium state or become inaccessible to the observables used to probe it. Thus, classicality depends on the interplay between the property, the system dynamics, and the accessible observables. In this respect, the result mirrors the role of the pointer basis in decoherence theory \cite{zurekRMP}, while requiring no external environment.

We illustrate these two mechanisms with a numerical simulation of a nonintegrable Ising chain in external fields, shown in Fig. \ref{fig:bounds}. We compare the exact time-averaged fluctuation with the bounds of Results 1 and 2. In the example shown, the equilibrium state retains coherence in the $B$ basis, so Result 1 does not provide a small bound. Nevertheless, the coherence component of the observable that survives equilibration is strongly suppressed, yielding a small bound from Result 2. Thus, the property $B$ becomes operationally classical even though coherence in the $B$ basis remains present in the equilibrium state. In particular, this occurs for a property $B$ that does not approximately commute with $H$, demonstrating that the observable-based mechanism is not restricted to approximately conserved properties.
\newline

\noindent\textit{Related Works:} 
Recently, connections between quantum equilibration and the emergence of classicality have been proposed in different contexts \cite{schwarzhans2025, engineer2024, demelo2025, Carvalho26, meier2025emergence}\footnote{Previous works have also argued that even in the absence of an external environment, inaccessible internal degrees of freedom can induce self-induced decoherence \cite{castagnino2013,castagnino2008,castagnino2000}. This mechanism has been compared to environmental decoherence in \cite{Schlosshauer05}; its precise relationship to modern equilibration results for isolated quantum systems remains to be clarified.}. Our viewpoint is in the same spirit, but generalizes the equilibration bound to classicality and emphasizes an operational notion of classicality: indistinguishability from a classical mixture when one probes the system with a restricted family of observables.

In Ref.~\cite{schwarzhans2025}, the authors identify conditions under which the equilibrium state $\rho_H$ exhibits the structural features expected from the quantum Darwinism program (in particular, a spectrum-broadcast form), implying redundant encoding of information about the system state in many fragments of an environment. Their results are complementary to ours, as they focus on classicality at the level of the equilibrium state and its information-theoretic structure.

In Ref.~\cite{demelo2025}, equilibration arguments are combined with random-matrix considerations in a measurement-model setting in which there is limited control over the Hamiltonian. The analysis yields conditions under which a classical limit emerges from the interaction between the system and the measuring device. While both \cite{demelo2025, schwarzhans2025} adopt an explicit system--apparatus/environment setup, here we instead ask when the system's own dynamics make its coherence operationally inaccessible to broad classes of probes, without committing to a specific measurement interaction.

Finally, we should note that operational approaches to classicality that account for finite resources are not new. In particular, many works show that coarse-graining alone is sufficient for the emergence of classical behavior; see, for example, \cite{KoflerBrukner2007, KoflerBrukner2008, Isadora20, Bibak2026}. In these works, however, no connection is made to the theory of equilibration in closed quantum systems.
\newline

\noindent\textit{Conclusions:} The central message of our work is that the emergence of classical behavior does not necessarily require the disappearance of quantum coherence. While Result 1 describes a decoherence-like mechanism in which the equilibrium state itself becomes approximately classical, Result 2 shows that classical behavior may arise even when substantial coherence remains present. In this case, equilibration renders the surviving coherence operationally inaccessible to the observables used to probe the system. This suppression of coherence can occur for physical properties that do not commute with the Hamiltonian, as illustrated by our numerical results, demonstrating that operational classicality need not be confined to approximately conserved properties.

Our results also reveal a close conceptual connection between equilibration and decoherence. In both cases, classical behavior emerges through the loss of operational access to coherence. However, unlike conventional decoherence, the mechanisms identified here do not rely on coupling to an external environment and apply entirely within isolated quantum systems.

\textit{Acknowledgments}
The authors thank Maximilian Lock and Tom Rivlin  for
fruitful discussions.
This work is supported in part by the National Council for Scientific and Technological Development, 
CNPq Brazil (projects: Universal Grant No. 408990/2025-2, and projects  409611/2022-0), 
and it is part of the National Institute of Science and Technology for Applied Quantum Computing (INCT-CQA) through CNPq process No. 408884/2024-0. 
TRO acknowledges funding from the Air Force Office of Scientific Research under Grant No. FO9550-23-1-0092. 
TD acknowledges financial support from CNPq (Grants No. 402786/2023-8 and No. 445150/2024-6, 310854/2026-1).
FdM acknowledges financial support from CNPq (Grant No. 305071/2022-0), and from  the Carlos Chagas Foundation for Research Support of the
State of Rio de Janeiro (FAPERJ, Grant APQ1 E-
26/210.576/2024)

\bibliography{ref} 

\appendix

\section{Proofs of Results 1 and 2} 

The general scheme to prove both results is as follows:  consider any observable $O$ and
\beq\label{eq:target_bound}
\overline{\big( \tr[\hat{O}(\rho_t-\rho_B)]\big)^2}.
\eeq


Now, still in general, insert and subtract the energy-dephased state $\rho_H$ in \eqref{eq:target_bound},
\beq
\tr[\hat{O}(\rho_t-\rho_B)] = \tr[\hat{O}(\rho_t-\rho_H)] + \tr[\hat{O}(\rho_H-\rho_B)].
\eeq
Denote
\beq
A(t)\equiv \tr[\hat{O}(\rho_t-\rho_H)],\qquad C\equiv \tr[\hat{O}(\rho_H-\rho_B)].
\eeq
Since $C$ is time independent, we have the identity
\beq\label{eq:split_bound}
\overline{(A(t)+C)^2} = \overline{A(t)^2} + C^2 + 2C\,\overline{A(t)}.
\eeq
Moreover, $\overline{A(t)}=\tr[\hat{O}(\overline{\rho_t}-\rho_H)]=0$ (because $\overline{\rho_t}=\rho_H$ under the same assumptions used to derive Eq.~\eqref{Eq. rho_average}). Therefore,
\begin{align}
\label{eq:split_bound_simplified}
\overline{\big( \tr[\hat{O}(\rho_t-\rho_B)]\big)^2}
=&\overline{\big( \tr[\hat{O}(\rho_t-\rho_H)]\big)^2}+\nonumber \\
&+\big( \tr[\hat{O}(\rho_H-\rho_B)]\big)^2.
\end{align}
The first term is controlled by the equilibration bound,
\beq
\overline{\big( \tr[\hat{O}(\rho_t-\rho_H)]\big)^2}\;\le\;\frac{\|\hat{O}\|_\infty^2}{d_{\rm eff}}.
\eeq

\subsection{Result 1.}
Now, specializing for the first results, for the second term in \eqref{eq:split_bound_simplified}, Hölder’s inequality  directly gives
\beq
\Big|\tr[ \hat{O} (\rho_H - \rho_B) ]\Big|^2\le\;\|\hat{O}\|_\infty^2 \| \rho_H - \rho_B\|_1^2.
\eeq
Choosing, if desired, $\rho_B=C_B(\rho_H)$, we reach to the statement of Result 1.

\subsection{Result 2.}

For the second result, we explicitly use that $\rho_B=C_B(\rho_H)$. Then the second term of \eqref{eq:split_bound_simplified} becomes
\begin{align}
\big| \tr[\hat{O}(\rho_H-\mathcal{C}_B(\rho_H)]\big|^2 
& = \big| \tr[\rho_H(\hat{O} -\mathcal{C}_B(\hat{O}))]\big|^2 \\
& = \big| \tr[C_H(\rho_H)(\hat{O} -\mathcal{C}_B(\hat{O}))]\big|^2 \\
& = \big| \tr[\rho_H \mathcal{C}_H(\hat{O} -\mathcal{C}_B(\hat{O}))]\big|^2 \\
& = \big| \tr[\rho_H \mathcal{C}_H(\hat{O}_B^{\rm coh})]\big|^2 \\
& =\big( \sum_n p_n \<E_n|\mathcal{C}_H(\hat{O}_B^{\rm coh})|E_n\>\big)^2 \\
& \leq \sum_n p_n \big( \<E_n|\mathcal{C}_H(\hat{O}_B^{\rm coh})|E_n\>\big)^2\\
& \leq \| \mathcal{C}_H(\hat{O}_B^{\rm coh})\|_\infty^2
\end{align}
In the first line and third line, we used the self-adjointness of the dephasing map; in the second line we used that the equilibrium state is already diagonal in the energy eigenbasis; in the fifth line, we only took the trace in the $H$ basis; and in the sixth line, we used the convexity of the square function. Finally, we used that $\<\psi|A|\psi\>^2 \le \|A\|_\infty^2$ for any operator $A$ and that $\sum_n p_n=1$.

\end{document}